# Creating a Systematic ESG (Environmental Social Governance) Scoring System Using Social Network Analysis and Machine Learning for More Sustainable Company Practices

Aarav Patel [1] and Peter Gloor [2]*

[1] Amity Regional High School – email: aarav.dhp@gmail.com

[2] Center for Collective Intelligence, Massachusetts Institute of Technology – email: pgloor@mit.edu

* Corresponding author.

## Abstract

Environmental Social Governance (ESG) is a widely used metric that measures the sustainability of a company's practices. Currently, ESG is determined using self-reported corporate filings, which allows companies to portray themselves in an artificially positive light. As a result, ESG evaluation is subjective and inconsistent across raters, giving executives mixed signals on what to improve. This project aims to create a data-driven ESG evaluation system that can provide better guidance and more systemized scores by incorporating social sentiment. Social sentiment allows for more balanced perspectives which directly highlight public opinion, helping companies create more focused and impactful initiatives. To build this, Python web scrapers were developed to collect data from Wikipedia, Twitter, LinkedIn, and Google News for the S&P 500 companies. Data was then cleaned and passed through NLP algorithms to obtain sentiment scores for ESG subcategories. Using these features, machine-learning algorithms were trained and calibrated to S&P Global ESG Ratings to test their predictive capabilities. The Random-Forest model was the strongest model with a mean absolute error of 13.4% and a correlation of 26.1% (p-value 0.0372), showing encouraging results. Overall, measuring ESG social sentiment across sub-categories can help executives focus efforts on areas people care about most. Furthermore, this data-driven methodology can provide ratings for companies without coverage, allowing more socially responsible firms to thrive.

# 1. Introduction

Many feel companies need to place more emphasis on social responsibility. For instance, 100 companies have been responsible for 71% of global greenhouse gas emissions since 1998 (Carbon Majors Database[1]). Many business leaders have publicly stated that they are on board with incorporating sustainability measures. In 2016, a UN survey found that 78% of CEO respondents believed corporate efforts should contribute to the UN Standard Development Goals, which are goals adopted by the United Nations as a universal call to action to end poverty and protect the planet (UN, 2016). However, while many executives pledged greater focus on these areas of concern, only a few took noticeable tangible action. In a more recent 2019 UN survey, only ~20% of responding CEOs felt that businesses were making a difference in the worldwide sustainability agenda (UN, 2019). These surveys highlight a disconnect between sustainability goals and sustainability actions. They also highlight inefficiencies in current executive actions since many feel they are not making enough progress toward social responsibility.

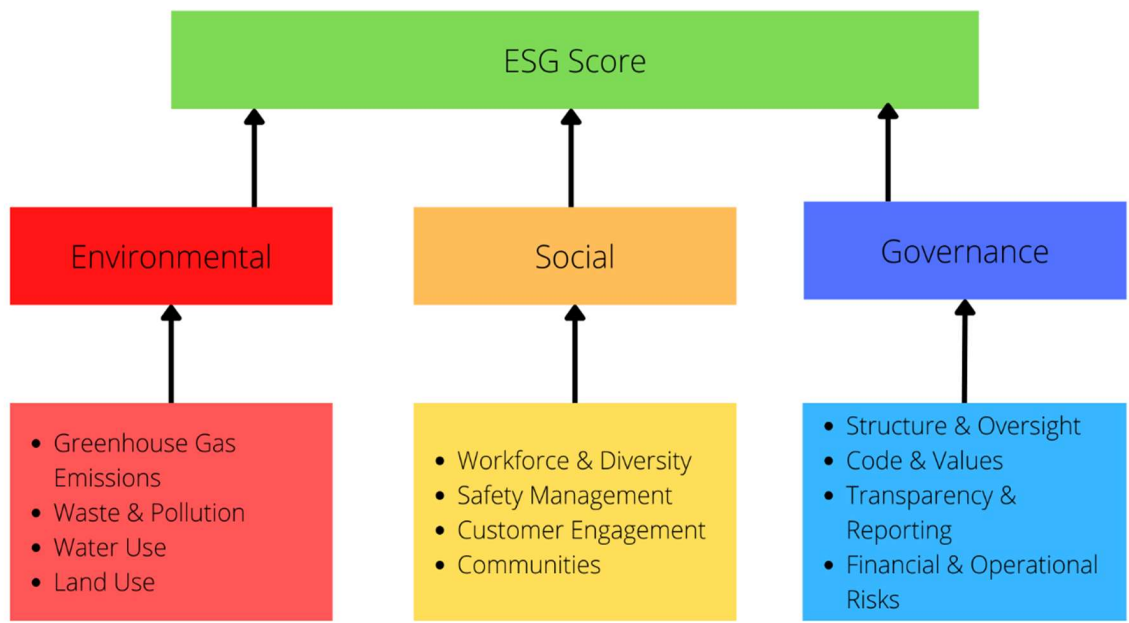

**Figure 1:** *Drawing inspired by S&P Global ESG evaluation framework (S&P Global)*

ESG, or Environmental Social Governance, is a commonly used metric that determines the sustainability and societal impact of a company's practices. ESG raters such as MSCI (Morgan Stanley Capital International), S&P Global, and FTSE (Financial Times Stock Exchange) do this by measuring sub-categories such as pollution, diversity, human rights, community impact, etc. (figure 1). Measuring these areas of concern are necessary since they encourage companies to rectify bad practices. This is because ESG ratings can influence factors such as investor capital, public perception, credit ratings, etc. Furthermore, ESG ratings can provide companies with specific information on which key areas to improve, which can help better guide their initiatives.

At the moment, ESG is assessed by rating agencies using self-reporting company filings. As a result, companies can often portray themselves in an artificially positive light. These biased reports have led to subjective and inconsistent analysis between different ESG rating organizations, despite them seeking to measure the same thing (Kotsanonis et al., 2019). For instance, the correlation among six prominent ESG rating agencies is 0.54; in comparison, mainstream credit ratings have a stronger correlation of 0.99 (Berg et al., 2019). As a result, many feel there is a disconnect between ESG ratings and a company's true social responsibility. This highlights how subjective assessment and limited data transparency from self-reporting can create inconsistent ratings.

Having more consistent and accurate ESG evaluation is important. Divergence and imprecision in ESG ratings hamper motivation for companies to improve since they give executives mixed signals on what to change (Stackpole, 2021). As a result, it becomes difficult to create better-targeted sustainability initiatives. Furthermore, self-reporting allows companies with more resources to portray themselves better. This is why there is a significant positive correlation between a company's size, available resources, and ESG score (Drempetic et al., 2019). These issues ultimately defeat the purpose of ESG by failing to motivate companies toward sustainable practices. This raises the need for a more holistic and systemized approach to ESG evaluation that can more precisely measure a company's social responsibility. By establishing a more representative ground truth, it can better guide company initiatives toward social responsibility, thus increasing the impact of ESG.

## 2. Related Works

Existing ESG-related research falls under two main categories. Some papers aim to correlate ESG performance with financial performance and see if a company's Corporate Social Responsibility (CSR) can be used to predict future stock performance (Jain et al., 2019). Other papers propose new data-driven methods for enhancing and automating ESG rating measurement to avoid existing fallacies/inefficiencies (Hisano et al., 2020; Krappel et al., 2021; Liao et al., 2017; Lin et al., 2018; Shahi et al., 2011; Sokolov et al., 2021; Venturelli et al., 2017; Wicher et al., 2019). This paper will fall into the latter category.

Since many firms publish sustainability reports on an annual basis, many researchers use this content for analysis. This is typically done using text mining to identify ESG topics and trends. In order to parse out and leverage this data, researchers have created classification models that can classify sentences/paragraphs into various ESG subdimensions (Liao et al., 2017; Lin et al., 2018). Additionally, some researchers have used these text classification algorithms to analyze the completeness of sustainability reports (Shahi et al., 2011). This is because companies sometimes limit disclosure regarding negative ESG aspects within their filings. Both tools can assist in automatic ESG scoring using company filings, which increases access for companies without ESG coverage.

However, there are deficiencies in solely relying on self-reported filings for analysis since it fails to consider omitted data or newer developments. As a result, researchers have been testing out alternative methods to solve this. For instance, some researchers utilize Fuzzy Expert System (FES) or a Fuzzy Analytic Network Process (FANP), pulling data from quantitative indicators (i.e., metrics provided by the Global Reporting Initiative) and qualitative features from surveys/interviews (Venturelli et al., 2017; Wicher et al., 2019). Others collected data from online social networks like Twitter to analyze a company's sustainability profile. For example, some used Natural Language Processing (NLP) frameworks to classify Tweets into various ESG topics and determine whether they are positive or negative (Sokolov et al., 2021). Furthermore, some used heterogeneous information networks that combined data from various negative news datasets and used machine learning to predict ESG (Hisano et al., 2020). Finally, others explored the viability of using fundamental data such as a company's profile and financials to predict ESG (Krappel et al., 2021). Overall, all these methods aimed to improve self-reported filings by using more balanced, unbiased, and real-time data.

## 3. Purpose

The purpose of this project was to create a systematic ESG rating system that gives executives and outsiders a more balanced and representative view of a company's practices for greater social responsibility. To do this, a machine-learning algorithm was created using social network data to quantitatively evaluate ESG. Social network data was used instead of self-reported filings since it can provide various outsider perspectives on issues people feel a corporation should address. By directly showcasing public opinion, it can remove the bias of self-reporting and help executives create more targeted initiatives for meaningful change. Furthermore, a data-driven system can provide ESG ratings for companies without coverage.

To test the predictive power of the proposed system, the correlation as well as the mean absolute average error (MAAE) were measured against current ESG ratings. This can help determine whether the system is viable for rating prediction. However, potential constraints include limited access to high volumes of social network data, the accuracy of NLP algorithms, and limited computational resources.

The contributions of this work can be summarized as follows:
- It gives a real-time social-sentiment ESG score that highlights how people feel regarding a company's practices. This can give executives a way to monitor the ESG health of their organization. It also shows which areas the people feel need the most change, and this can help target executive initiatives to be more effective.
- It provides a full-stack method for gathering real-time ESG data and converting it into a comprehensive score. This allows for the readily available creation of initial ESG ratings that can be used either directly by investors to ensure they are making socially conscious investments (especially for non-rated companies) or by ESG rating agencies to scale up coverage.
- The proposed approach utilizes multiple social networks for score prediction. Most papers about ESG social network analysis typically hyperfocus on one specific network such as Twitter or the News (Sokolov et al., 2021). This paper seeks to combine them while also adding other under-analyzed social networks (i.e., LinkedIn, Wikipedia).

## 4. Methods

The creation of this project was divided into three steps. The first step was data collection through web scrapers across various social networks. Afterward, text data was pre-processed and converted into sub-category scores using Natural Language Processing. Finally, machine-learning algorithms were trained using this data to compute a cohesive ESG rating.

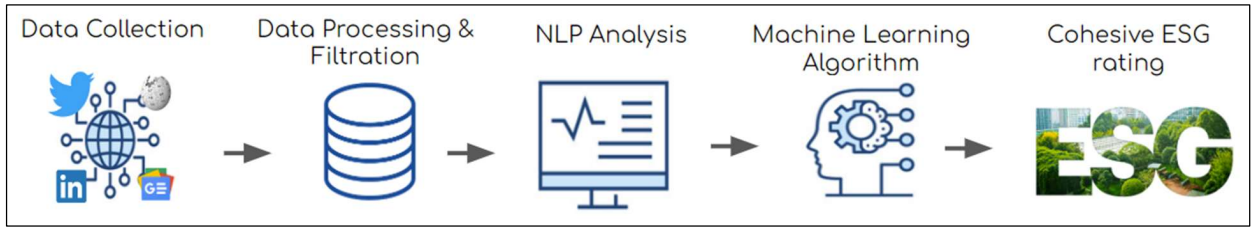

**Figure 2:** *An overview of how the data-driven ESG index uses social network data to compute a cohesive ESG rating*

### 4.1. Data Collection

Rather than use self-reported corporate filings, social network data was used to holistically quantify ESG. Social network analysis and web scraping can be used to identify trends (Gloor et al., 2009). Popular social networks such as Twitter, LinkedIn, and Google News have a plethora of data pertaining to nearly any topic. This data can provide a balanced view of company ESG practices, and it can help cover both short-term and long-term company ESG trends. It can also gather data that might not be reflected in filings. Finally, this data can directly highlight the concerns of outsiders, which can better guide company ESG initiatives to be more impactful.

**Environment:** environment, carbon, climate, emission, pollution, sustainability

**Social:** social, community, discrimination, diversity, human rights, labor

**Governance:** governance, compensation, corruption, ethical, fraud, justice, transparency

**Figure 3:** *Keywords/Topics used for data collection*

To do this, a comprehensive list of ESG-relevant keywords was created (figure 3). This list of keywords was inspired by sub-categories commonly used in current ESG rating methodologies. This list was used to help collect publicly available company data from Wikipedia, LinkedIn, Twitter, and Google News. To collect data, web scrapers were developed in Python. Wikipedia data was collected using the Wikipedia Application Programming Interface (API). Wikipedia serves to give a general overview of a company's practices. Google News data was collected by identifying top news articles based on a google search. The links to these articles were stored. The news serves to give overall updates on notable ESG developments. Twitter data was collected with the help of the Snscrape library. Snscrape is a lightweight API that allows users to collect near unlimited Tweets (with certain restrictions on how many can be collected per hour) from almost any timeframe. Twitter was chosen to primarily give consumer-sided feedback on a company's practices. Since the LinkedIn API does not support the collection of LinkedIn posts, an algorithm was created from scratch to do so instead. The algorithm utilized the Selenium Chromedriver to simulate a human scrolling through a LinkedIn query. Based on this, each post's text was collected and stored using HTML requests via BeautifulSoup. LinkedIn serves to provide more professional-sided information on a company's practices. This data collection architecture allows for ratings to be refreshed and generated in real time as needed. Afterward, data for each sub-category was stored in a CSV file.

These four social networks cover a wide range of company ESG data. Data was collected for most S&P 500 companies (excluding real estate). Real estate was excluded primarily because it did not receive as much coverage pertaining to ESG issues (based on surface-level analysis), so it did not seem viable for the proposed system. This ensures the collected companies were well balanced across sectors and industries. The web scrapers attempted to collect ~100 posts/articles for each keyword on a social network. However, sometimes less data would be collected because of API rate limits and limited data availability for the lesser-known companies. In order to speed up collection, multiple scripts were run simultaneously. At first, the programs would often get rate-limited for collecting so much data in such a short timeframe. To resolve this, safeguards were added to pause the program in case it encountered this. All data collection was done following each site's terms and conditions. In total, approximately ~937,400 total data points were collected across ~470 companies, with an average of ~37 points per social network keyword. Most of this data was concentrated in 2021. However, a hard date range was not imposed because it would remove data points for lesser-known companies that already struggled to gather enough information.

Once all data was collected, it was exported onto a spreadsheet for further analysis. Data was preprocessed using RegEx (Regular Expressions). First, URLs and links were removed. Mentions were replaced with a generic word to abstractify names. Finally, Uncommon characters and punctuation were removed. This helped filter out words/characters that might interfere with NLP analysis.

**4.2. NLP Analysis**

After the data was cleaned and organized, an NLP algorithm was built for analysis. Firstly, an ESG relevancy algorithm was created to filter out ESG irrelevant data that might obstruct results. To do this, keyword detection was used to see if the post/article discussed the current company as well as one or more of the ESG sub-categories. Next, Python's Natural Language Toolkit (NLTK) Named Entity Recognition library was used to determine if a post related to the organization in order to remove unintended data. For example, if the query "apple climate" was searched, then a post might come up saying "Spring climate is the best time to grow apple trees." However, Named Entity Recognition would be able to identify that this sentence is not ESG relevant since "Apple" is used as an adjective. Therefore, the algorithm will disregard it from the analysis. On the other hand, if the post said, "Apple is pouring 500 million dollars into initiatives for climate change," then the algorithm would determine that the post is talking about Apple the organization. This filtration step helps remove irrelevant information to improve data quality.

After filtration, NLP sentiment analysis was used to score whether a post was ESG positive or negative. Two NLP algorithms were created to do this: the short-post NLP algorithm analyzed shorter bodies of text (Tweets, LinkedIn posts) while the long-article NLP algorithm analyzed longer ones (News articles, Wikipedia articles).

A literary analysis of different Python sentiment analysis libraries was carried out. After comparing various sentiment analysis libraries such as TextBlob, VADER, FastText, and Flair, it was found that Flair outperformed the other classifiers. This is likely because the simple bag-of-words classifiers, such as VADER or TextBlob, failed to identify the relations that different words had with each other. On the other hand, Flair used contextual word vectors to analyze a sentence's word-level and character-level relationships. This is likely why, when these algorithms were tested on the Stanford Sentiment Treebank (SST) to rate movie review sentiment on a scale of 1-5, it was found that the Flair algorithm performed the best with an F1 score of 49.90% (Akbik et al., 2018) (Rao et al., 2019) (figure 4). So, the short-post algorithm was built using the Flair sentiment analysis library. The long-article algorithm is essentially the short-post algorithm but averaged across all relevant body paragraphs (i.e., paragraphs containing the company name) in an article.

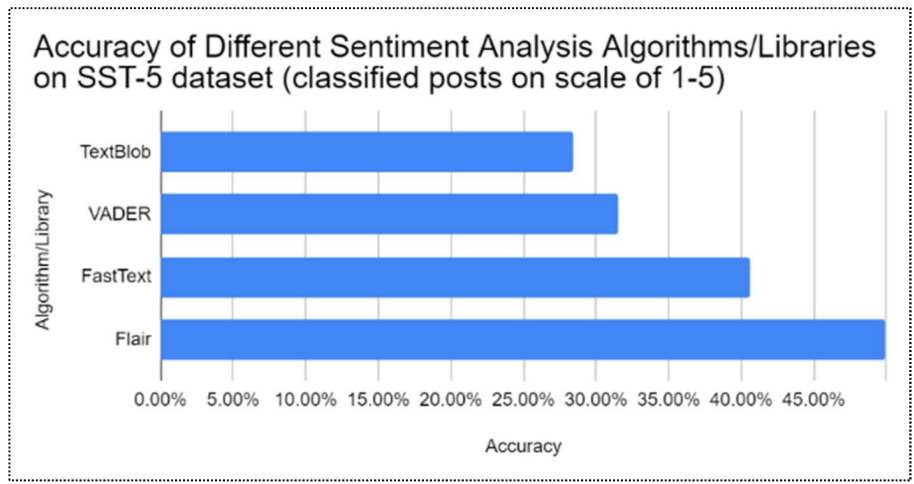

**Figure 4:** *Comparison of accuracy of different sentiment analysis algorithms on SST-5 database*

These umbrella algorithms were further optimized for each specific social network. For example, the LinkedIn algorithm analyzed the author's profile of a LinkedIn post to eliminate self-reporting. This is because executives often discuss their positive initiatives and goals, which can dilute other unbiased observations and thus construe results. Additionally, for the Twitter and LinkedIn algorithms, if a link address was found within the text, then the algorithm would analyze that article for evaluation.

Initially, the analysis algorithm was very slow since it would take Flair 3-4 seconds to analyze one post. So, a variation called "Flair sentiment-fast" was installed. This allowed Flair to conduct batch analysis where it analyzes multiple posts simultaneously. This significantly cut down on analysis time while slightly sacrificing accuracy.

Once all raw data was scored, the scores were averaged into a cohesive spreadsheet. Mean imputing was used to fill in any missing sub-score data. These sub-category scores can provide executives with breakdowns of social sentiment on key issues, giving them concrete information about which areas to improve. These scores can be used raw to help guide initiatives, or they can be compiled further through machine learning to provide an ESG prediction.

### 4.3. Machine Learning Algorithms

After compiling the data, different machine-learning models were tested. The goal of these models was to predict an ESG score from 0-100, with 0 being the worst and 100 being the best. Most of these supervised learning models were lightweight regression algorithms that can learn non-linear patterns with limited data. Some of these algorithms include Random Forest Regression, Support Vector Regression, K-Nearest Neighbors Regression, and XGBoost (Extreme Gradient Boosting) Regression. Random Forest Regression operates by constructing several decision trees during training time and outputting the mean prediction (Tin Kam Ho, 1995). Support Vector Regression identifies the best fit line within a threshold of values (Awad et al., 2015). K-Nearest Neighbors Regression predicts a value based on the average value of its neighboring data points (Kramer, 2013). XGBoost (Extreme Gradient Boosting) Regression uses gradient boosting by combining the estimates/predictions of simpler regression trees (Chen et al., 2016).

These regression algorithms were trained using 19 features. These features include the average sentiment for each of the 18 keywords with an additional category for Wikipedia. They were calibrated to public S&P Global ESG ratings to ensure they did not diverge much from existing solutions. A publicly licensed ESG rating scraper on GitHub was used to retrieve S&P Global ESG scores for all companies that were analyzed (Shweta-29). Optimization techniques such as regularization were used to prevent overfitting for greater accuracy.

Before creating the algorithms, companies with less than 5 articles/posts per ESG subcategory were filtered out. This left ~320 companies for analysis. In order to create and test the algorithm, ~256 companies were used as training data, while ~64 companies were used for testing data. These results were used to determine the predictive capabilities of the algorithm.

## 5. Results

The Random Forest Regression model displayed the strongest overall results when tested on a holdout sample of 64 companies. The Random Forest Regression model had the strongest correlation with current S&P Global ESG scores with a statistically significant correlation coefficient of 26.1% and a mean absolute average error (MAAE) of 13.4% (Figure 5, 6). This means that the algorithm has a p-value of 0.0372 (<0.05), showing that it is well-calibrated to existing ESG solutions. On the other hand, while the other models have similar MAAE, they also have lower correlation coefficients that do not prove to be statistically significant (Figure 6). For example, Support Vector Regression algorithm had a correlation of 18.3% and MAAE of 13.7%, which results in a p-value of 0.148 (Figure 8). The XGBoost model had a correlation of 16.0% and MAAE of 14.7%, which results in a p-value of 0.207 (Figure 7). Finally, the K-Nearest Neighbors algorithm had a correlation of 13.2% and a MAAE of 14.0%, which is a p-value of 0.298 (Figure 9). However, all the algorithms had a similar MAAE that fell between 13%-15%, with the Random Forest model having the lowest at 13.4% (Figure 10). All the algorithms surpassed the MAAE criteria of 20.0%.

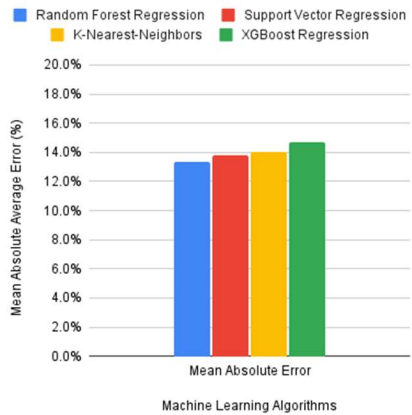

**Figure 5:** *Mean Absolute Average Error of different machine-learning algorithms against S&P Global ESG score*

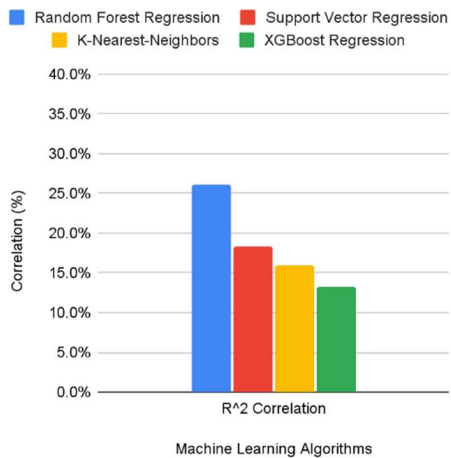

**Figure 6:** *$R^2$ correlation of different machine-learning algorithms*

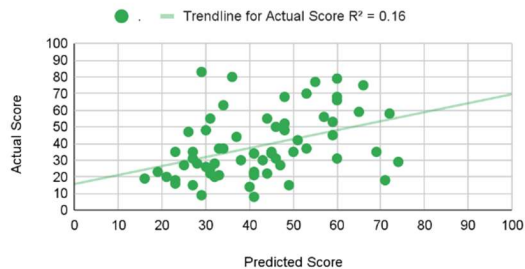

**Figure 7:** *XGBoost model predictions v actual scores (scale 0-100)*

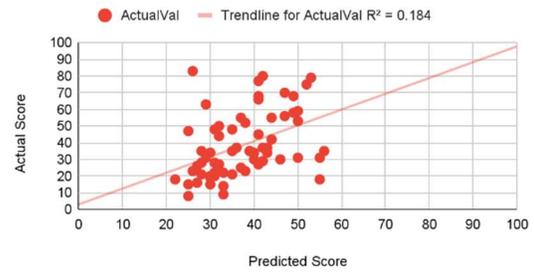

**Figure 8:** *Support Vector Regression predictions v actual scores (scale 0-100)*

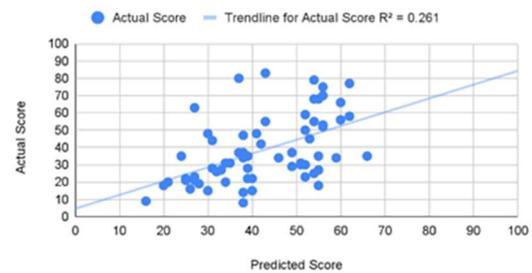

**Figure 9:** *K-Nearest Neighbor model predictions v actual scores (scale 0-100)*

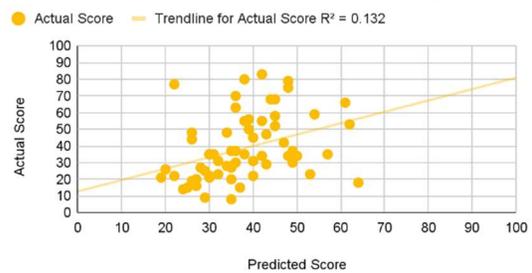

**Figure 10:** *Random Forest model predictions v actual scores (scale 0-100)*

# 6. Discussion

The Random Forest Regression Model likely performed the best because it works by combining the predictions of multiple decision trees. This allows it to improve its accuracy and reduce overfitting to one specific tree, thus producing superior results. The Random Forest Regression algorithm had a statistically significant $R^2$ Correlation of 26.1% (p-value <0.05), and it had a low MAAE of 13.4%. These results align with similar work done using other sources of data (Krappel et al., 2021). For example, a paper by Krappel et al. created an ESG prediction system by feeding fundamental data (i.e., financial data and general information surrounding the company) into ensemble machine-learning algorithms. Their most accurate model received an $R^2$ correlation of 54% and a MAAE of 11.3%. While the proposed algorithm does not correlate as well as Krappel et al.'s model, likely because it leverages qualitative data, it still highlights the viability of using social sentiment as a proxy for ESG.

The proposed algorithm displayed encouraging results, highlighting its viability in ESG rating prediction. Unlike current ESG raters who determine ESG using self-disclosed sustainability reports, the proposed algorithm's data-driven approach allows for a more holistic and balanced evaluation. Utilizing social sentiment also allows executives to measure which areas people want a company to improve on, helping to focus actions on change Additionally, the system's architecture allows for scores to be updated within short timeframes. Finally, executives can test additional keywords by inputting them into the algorithm. These attributes showcase the system's flexibility as well as advantages over the conventional methodology.

A limitation of the results, however, is that it was tested on the S&P 500 companies. Therefore, results might not carry over for smaller companies below this index. Another limitation could be misinformation within the social network data. While this should be diluted by other comments, it can potentially alter the algorithm's ratings. Additionally, the Flair sentiment analysis algorithm sometimes misclassified post/article sentiment, especially if the post/article had a sarcastic attitude. Finally, for this research, access to certain paid native APIs was not available. As a result, the collected data might not encompass all data available for a keyword due to rate limiting.

While the algorithm has displayed statistically significant results, there is room for improvement that can be done in future research. Some of this can include gathering more data. This can be done by analyzing more companies beyond the S&P 500 or by collecting data for more keywords and ESG sub-topics. This can also be done by using native APIs to collect more datapoints per individual keyword. Additionally, more data sources could be incorporated into the model. This can be done by incorporating other social networks (i.e., Reddit, Glassdoor) or by including quantitative data/statistics (i.e., % women as board members, number of scope 1 carbon emissions, etc.) from company reports and government databases.

Furthermore, to better fit the task at hand, NLP algorithms can be created specifically for ESG. For instance, while the current method filters much of the irrelevant data, some unrelated data still gets through. So, to solve this, a new supervised learning algorithm can be trained to identify related bodies of text using TF-IDF vectorization. The algorithm can be trained by hand-labeling the data that has already been collected. To add on, the long-article/short-post NLP algorithms can also be further optimized. While Flair can already provide satisfactory results, some articles seem to be misclassified, which might be a source of error for the algorithm. By creating a sentiment analysis algorithm specifically tailored to ESG classification, the long-article and short-post NLP algorithm accuracy can be further improved. This can be done by either creating a custom ESG lexicon with weights or by training a novel NLP algorithm against classified ESG data.

Finally, another area to be improved is post credibility: While small amounts of misinformation would not significantly alter results, it is still best to mitigate this risk as much as possible. There is a growing body of literature that explores fake news identification on social networks. So, these approaches can potentially be used to identify fake posts/articles (de Beer et al., 2020). Also, adding "hard" quantitative data from company filings to the algorithm can be used as an added safeguard. Finally, the algorithm can prioritize more centralized/credible actors over others to yield safer outputs.

Overall, this research provides a proof-of-concept framework for a social-network-based ESG evaluation system. This work can serve as the backend logic for a social sentiment ESG product which can eventually be used by executives. While pre-packaged libraries were used for prototyping purposes, in future works, these aspects of the project can be optimized. Unlike existing frameworks that rely on self-reported company filings, the proposed models take on a more balanced view of the company's ESG positives and negatives. In general, this can help approach an ESG ground truth that can better influence company practices to be more sustainable.

## 7. Conclusion

The proposed ESG analysis algorithm can help standardize ESG evaluation for all companies. This is because it limits self-reporting bias by incorporating outside social network analysis for more balanced results. A social-network-based ESG index can also directly show which areas people want to change, which can better focus executive efforts on meaningful change. Additionally, using machine learning, the model can generate a proxy for a company's social responsibility, which can help determine ESG for smaller companies that do not have analyst coverage. This will help more companies receive ESG ratings in an automated way, which can create a more level playing field between small and large companies and ultimately help more socially responsible firms prevail. Overall, the project can have broad implications for bridging the gap in ESG. This will help rewire large quantities of ESG capital to more sustainable and ethical initiatives.